\documentclass[9pt]{article}
\usepackage{microtype}      
\usepackage{amsmath,amssymb}
\usepackage{amsfonts} %
\usepackage{amsthm,bm}
\usepackage{mathtools}
\usepackage{subfigure}
\usepackage{graphicx}
\usepackage{comment}
\usepackage{xspace}
\usepackage{authblk}
\usepackage{ulem}
\usepackage{cite}
\usepackage{pdfpages}

\newtheorem{theorem}{Theorem}

\newtheorem{proposition}[theorem]{Proposition}

\newcommand{\fixme}[1]{\textbf{\color{red}[#1]}}

\newcommand{\figref}[1]{Fig.~\ref{fig:#1}}

\newcommand{\secref}[1]{Section~\ref{sec:#1}}

\newcommand{\eq}[1]{\eqref{eq:#1}}

\newcommand{\mA}{\mathbf{A}}
\newcommand{\mB}{\mathbf{B}}
\newcommand{\mC}{\mathbf{C}}
\newcommand{\mD}{\mathbf{D}}
\newcommand{\mG}{\mathbf{G}}

\newcommand{\mP}{\mathbf{P}}
\newcommand{\mQ}{\mathbf{Q}}
\newcommand{\mR}{\mathbf{R}}
\newcommand{\mS}{\mathbf{S}}
\newcommand{\mT}{\mathbf{T}}

\newcommand{\mV}{\mathbf{V}}
\newcommand{\mW}{\mathbf{W}}
\newcommand{\mX}{\mathbf{X}}
\newcommand{\mY}{\mathbf{Y}}
\newcommand{\mI}{\mathbf{I}}

\newcommand{\ma}{\mathbf{a}}
\newcommand{\me}{\mathbf{e}}
\newcommand{\mh}{\mathbf{h}}
\newcommand{\mb}{\mathbf{b}}

\newcommand{\mLam}{\bm{\Lambda}}

\newcommand{\um}[0]{$\mu\text{m}$\xspace}
\newcommand{\mum}[0]{\mu\text{m}}

\title{Differentiable Scattering Matrix for Optimization of Photonic Structures}
\author[1]{Ziwei Zhu}
\author[1,*]{Changxi Zheng}
\affil[1]{Department of Computer Science, Columbia University, New York, New York 10027, USA}

\affil[*]{Corresponding author: cxz@cs.columbia.edu}

\begin{document}
\maketitle

\begin{abstract}
    The scattering matrix, which quantifies the optical reflection and
    transmission of a photonic structure, is pivotal for understanding the
    performance of the structure.  In many photonic design tasks, it is also
    desired to know how the structure's optical performance changes with
    respect to design parameters, that is, the scattering matrix's derivatives (or gradient).
    Here we address this need. We present a new algorithm for computing
    scattering matrix derivatives accurately and robustly. In particular, we
    focus on the computation in semi-analytical methods (such as rigorous
    coupled-wave analysis).  To compute the scattering matrix of a structure,
    these methods must solve an eigen-decomposition problem.  However, when it
    comes to computing scattering matrix derivatives, differentiating the
    eigen-decomposition poses significant numerical difficulties.
    We show that the differentiation of the eigen-decomposition
    problem can be completely sidestepped, and thereby propose a robust
    algorithm.  To demonstrate its efficacy, we use our algorithm to optimize
    metasurface structures and reach various optical design goals.
\end{abstract}

\section{introduction}\label{sec:intro}
The scattering matrix is a fundamental concept in many fields.  It relates the
input state and the output state of a physical system undergoing a scattering
process. Particularly revealing in optics, the scattering matrix has been
widely used for analyzing photonic structures such as
waveguides~\cite{lalanne2000fourier,cao2002stable,kwiecien2015rcwa} and
metasurface units~\cite{kamali2017angle, zhao2018multichannel, blau2018situ}.
Once the scattering matrix of a photonic structure is known, the structure's
optical performance (e.g., mode conversion efficiency and phase shift) can be
directly obtained.

Because of its vital importance, 
many numerical methods have been developed to compute the scattering matrix
of a photonic structure. 
Among them, a popular class is the \textit{semi-analytical methods}, 
such as the method of
lines~\cite{pregla1989method} and rigorous coupled-wave analysis (RCWA)~\cite{moharam1981rigorous}.
These methods exploit
the fact that many photonic structures in practice (such as waveguides and metasurface units)
have a piecewise constant cross-sectional shape along the transmission direction (denoted as $z$-direction). 
Thus, to solve Maxwell's equations, they only need to discretize the 2D cross-sectional region, 
reducing Maxwell's equations into a set of
continuous differential equations along $z$-direction, whose solution can be
expressed through an eigenvalue analysis.
Thanks to the semi-discretization, these methods often enable 
faster computation in comparison to full discretization methods 
(such as finite-element- and finite-volume-based methods).
Indeed, methods like RCWA have been widely used in designing 
various photonic structures, such as
metasurfaces~\cite{arbabi2015dielectric,kamali2017angle,arbabi2016multiwavelength,colburn2018varifocal},
metagratings~\cite{fitio2017application, ahmed2010design},
holograms~\cite{zhao2018multichannel, zhou2019multifunctional},
polarimeters~\cite{arbabi2018full}, 
solar cells~\cite{wang2012absorption}, 
radiative cooling structures~\cite{rephaeli2013ultrabroadband}, 
color structures~\cite{shen2015structural}, photonic crystals\cite{wang2014spectral}, 
and waveguides~\cite{lalanne2000fourier, cao2002stable}.

In this work, we extend the {semi-analytical} methods 
to obtain the higher-order information of scattering matrices, namely the scattering matrix's \textit{derivatives}
(or gradient).  Provided a photonic structure specified by certain design
parameters, we aim to compute not only its scattering matrix but its
derivatives with respect to the design parameters. 

The scattering matrix derivatives depict the changes of the
structure's optical behaviors as its design parameters vary.
This higher-order information, 
if robustly and efficiently computed, 
finds many applications in photonic design. Most notable is the optimization of 
photonic structures. 
The derivatives provide guidance on how we can adjust the parameters (e.g., through 
the gradient descent algorithm~\cite{ruder2016overview}) to improve the structure's optical
performance~\cite{lalau2013adjoint} or to find a design robust to 
fabrication error~\cite{akel2000design, veronis2004method, ziwei2020}. 

Unfortunately, the computation of scattering matrix derivatives is nontrivial. 
The difficulty is rooted in the fact that the permissible optical modes in 
a photonic structure are eigenfunctions of a linear (Hermitian) operator
determined by Maxwell's equations~\cite{joannopoulos1995photonic}.
Thus, to compute the scattering matrix, semi-analytical methods must solve an
eigen-decomposition problem:  
its eigenvalues describe the propagation constants (or effective indices) of the modes
and its eigenvectors indicate propagating modal patterns.
Differentiating the scattering matrix, by chain rule,
requires the derivatives of eigenvalues and eigenvectors.
It is the need of eigenvector derivatives that renders the scattering matrix differentiation ill-posed:
when there exist repeated eigenvalues, 
the corresponding eigenvectors are not uniquely defined. 
As the parameter changes, the numerical results of the eigenvectors
may change discontinuously, and their derivatives become undefined 
(see more discussion in \secref{diff}). 

Not merely does this issue exist  as a corner case; many photonic structures
in practice have geometric and material symmetries, from which repeated
eigenvalues and thus ill-defined eigenvector derivatives emerge (see \figref{repeated_ev}).
Consequently, 
one must carefully choose eigenvectors such that they 
vary smoothly with respect to the design parameters.
This choice, albeit attainable, demands complex and expensive computational effort~\cite{van2007computation}.

In this paper, we question the necessity of eigenvector derivatives for
differentiating scattering matrices.
We show that while eigen-decompositions
are needed for computing a photonic structure's scattering matrix,
eigenvector derivatives can be fully sidestepped for differentiating the scattering 
matrix. 
Based on our new derivation, we present a fast and robust algorithm that,
without resorting to eigenvector derivatives, computes the scattering matrix
derivatives with respect to any design parameters.

Our method is designed for scattering matrices in general,
independent from any specific basis representation;
nor is it bound to any particular geometric parameterization.
To demonstrate the use of our method, we apply the scattering matrix
derivatives for optimizing the design of metasurface units.  
We can choose different design parameterizations and 
use gradient-based optimization to reach various light transmission goals.
We also propose a general parameterization of the meta-unit's cross-sectional shape
that can be optimized using our method.

\section{Background: Scattering Matrix}\label{sec:classic}
We start by briefly reviewing the classic notion of scattering matrix in computational photonics, 
to pave the way toward its differentiation.

\begin{figure}[t]
\centering
\includegraphics[width=\linewidth]{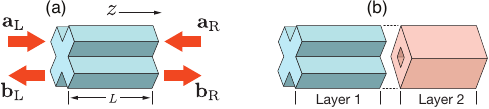}
\vspace{-1mm}
\caption{(a) In a photonic structure, light may be incident from both sides
    and get scattered out. The relationship of incident and scattered light is 
    characterized by the scattering matrix.
    (b) A complex structure can be decomposed into individual layers. Each layer is
    characterized by its scattering matrix, and these scattering matrices are then
    combined (using the Redheffer star product~\cite{redheffer1961difference}) to form the scattering matrix of the entire structure.
\label{fig:photonic_structure}}
\vspace{-3mm}
\end{figure}
To numerically analyze a photonic structure (such as a waveguide),
the structure is often discretized along the wave propagation direction (i.e., $z$-direction) 
into a series of layers each with a uniform cross-sectional material distribution (\figref{photonic_structure}-b).
Consider optical waves of a specific frequency.
Their propagation in each layer is characterized by a \textit{scattering matrix} $\mS$,
which relates waves incident on the layer from left and right sides (\figref{photonic_structure}-a) to the waves scattered out in either direction.

Concretely, let $\ma_{\text{L}}$ and $\ma_{\text{R}}$ denote vectors describing
incident waves on the layer from left and right sides, respectively.
Here $\ma_{\text{L}}$ and $\ma_{\text{R}}$ stack coefficients that represent the 
waves under a chosen basis, {whose construction will be outlined shortly.}
Under the same basis, we use $\mb_{\text{L}}$ and $\mb_{\text{R}}$  to denote 
the scattered waves in left and right directions. 
With these notations, the incident and scattered waves are related through
\begin{equation}\label{eq:scattering} 
    \begin{bmatrix}
        \mb_{\text{L}} \\
        \mb_{\text{R}}
    \end{bmatrix} = \mS 
    \begin{bmatrix}
        \ma_{\text{L}} \\
        \ma_{\text{R}}
    \end{bmatrix},
    \text{ where }
\mS \coloneqq \begin{bmatrix}
    \mR_{\text{L}} & \mT_{\text{RL}} \\
    \mT_{\text{LR}} & \mR_{\text{R}}
\end{bmatrix}.
\end{equation}
Here $\mS$ is decomposed into four submatrices: $\mR_{\text{L}}$ and
$\mR_{\text{R}}$ indicate how the incident wave from left or right direction 
is reflected by the layer, while $\mT_{\text{RL}}$ and $\mT_{\text{LR}}$
describe how the incident wave (from either direction) transmits through the layer.

\begin{figure}[t]
\centering
\includegraphics[width=\linewidth]{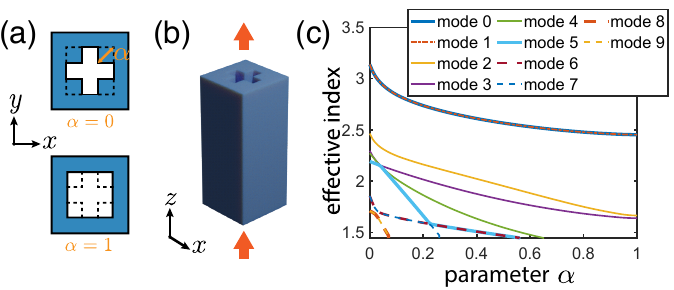}
\vspace{-3mm}
\caption{\textbf{Repetition from symmetry.} 
Consider a meta-atom structure (b)
whose cross-sectional shape (a) is parameterized by $\alpha$, 
which controls the size of the cross-shaped hollow region.
The shape symmetry causes the structure's propagating modes
to have repeated effective indices (c), as also indicated by the repeated eigenvalues
when one solves~\eq{pq_form}. As $\alpha$ varies, mode 0 and mode 1 always have the same
effective index, meaning that their first-order and higher-order derivatives of the 
corresponding eigenvalues with respect to $\alpha$ are always the same,
and thus their eigenvector derivatives are not mathematically well-defined.
The same issue occurs in other modes when $\alpha$ becomes small and more propagating modes appear
(e.g., see mode 5 and mode 6 when $\alpha$ is within $\sim[0.3,0.5]$).
\label{fig:repeated_ev}}
\vspace{-2mm}
\end{figure}

\textbf{The computation of scattering matrix} starts with a semi-discretization of the 
frequency-domain Maxwell’s equations of a photonic layer, namely,
\begin{equation}\label{eq:pq_form}
    -jk_0\frac{\partial}{\partial z}\me = \mP \mh
    \;\textrm{ and }\;
    -jk_0\frac{\partial}{\partial z}\mh = \mQ \me,
\end{equation}
where $k_0$ is the free-space wave number, and the vectors $\me$ and $\mh$ describe the electric and magnetic fields of the photonic structure
under a chosen basis---for example, RCWA uses the 2D Fourier basis on the 
cross-section of the wave propagation direction. 
The matrices $\mP$ and $\mQ$ encode the cross-sectional distributions of material permeability 
and permittivity.

The semi-discretization~\eq{pq_form} is a common form in many  
numerical analysis methods for photonic structures (such as the method of line~\cite{pregla1989method} and RCWA~\cite{moharam1981rigorous}).
The difference across those methods only lies in the specific ways of constructing $\mP$ and $\mQ$ 
(e.g., see {Supplement 1} for their construction in RCWA).

Once $\mP$ and $\mQ$ are determined, the scattering matrix $\mS$ can be constructed.
A key step of this construction is to solve an eigenvalue problem, 
$(\mP\mQ)\mW = \mW\bm{\Gamma}$, to obtain eigenvectors $\mW$ and the diagonal eigenvalue matrix
$\bm{\Gamma}$. As we will discuss in \secref{diff}, it is this eigenproblem 
that renders the differentiation of the scattering matrix ill-posed.
To understand the challenges and how we overcome them,
we first present the recipe of computing $\mS$ from $\mW$ and $\bm{\Gamma}$,
as follows. 

Let $\bm{\Omega}\coloneqq(\mP\mQ)^{\frac{1}{2}}$. Then, its eigenvalue matrix 
is $\mLam = \bm{\Gamma}^{\frac{1}{2}}$.
As derived in~\cite{rumpf2011improved}, the formulas of computing the scattering matrix $\mS$ defined 
in~\eq{scattering} are
\begin{subequations}\label{eq:s_formula}
  \begin{align}
    \mR_{\textrm{L}} = \mR_{\textrm{R}} & = \left(\mA -\mX\mB\mA^{-1}\mX\mB\right)^{-1}\left(\mX\mB\mA^{-1}\mX\mA-\mB\right),\label{eq:s11} \\
    \mT_{\text{LR}}=\mT_{\text{RL}} & = \left(\mA -\mX\mB\mA^{-1}\mX\mB\right)^{-1}\mX\left(\mA - \mB\mA^{-1}\mB\right), 
  \end{align}
\end{subequations}
where the matrices $\mX$, $\mA$, and $\mB$ have the following forms:
\begin{subequations}\label{eq:abx}
  \begin{align}
      \mX &= e^{j\mLam\frac{L}{k_0}},\label{eq:abx-a} \\
    \mA &= \mW^{-1}\mW_0 + \mV^{-1}\mV_0,\;
    \text{ and }
    \mB = \mW^{-1}\mW_0 - \mV^{-1}\mV_0.
  \end{align}
\end{subequations}
Here we use $L$ to denote the layer thickness (\figref{photonic_structure}), and the matrix $\mV$
is related to $\mW$ through $\mV=\mQ\mW\mLam^{-1}$.
$\mW$ and $\mV$ together form a basis of electric and magnetic components 
for the optical waves in the layer.
Similarly, $\mW_0$ and $\mV_0$ form a basis for free space propagation,
independent from the photonic structure. {They} are constant values
for computing the derivatives of $\mS$.
The vectors, $\ma_{\text{L}}$, $\ma_{\text{R}}$ $\mb_{\text{L}}$, and $\mb_{\text{R}}$,
in~\eq{scattering} are coefficients under this free-space basis 
to describe incident and scattered waves.

Once the scattering matrices of individual layers are computed, 
they are combined using the Redheffer star product~\cite{redheffer1961difference}
into the total scattering matrix, one that indicates the optical response of the entire
photonic structure.

\textit{Remark.} The formulas in Eqs.~(\ref{eq:s_formula}) and (\ref{eq:abx})
assume that the current photonic layer is sandwiched by two free-space layers.
This assumption is by no means a limitation. In an arbitrary photonic 
structure, the layers can be 
treated as if they are interleaved with free-space layers---each of which has a zero thickness.


\section{Differentiable Scattering Matrix}\label{sec:diff}
The geometry or material distribution of photonic structure is specified by its
structural (design) parameters (e.g., see \figref{repeated_ev}).  These
parameters determine the structure's permittivity and {permeability}
distributions described by $\mP$ and $\mQ$ in~\eq{pq_form}.  Thus, one can
compute their derivatives, $\mP'$ and $\mQ'$, with respect to an arbitrary
parameter.  Given $\mP'$ and $\mQ'$, we now address the question of how to
compute the scattering matrix derivative $\mS'$ with respect to the same
parameter.

\subsection{Challenges in Scattering Matrix Differentiation}
The construction of scattering matrix $\mS$ 
needs to solve an eigenvalue problem $(\mP\mQ)\mW = \mW\bm{\Gamma}$, 
as the eigenvalues $\bm{\Gamma}$ and eigenvectors $\mW$ appear in
Eqs.~(\ref{eq:s_formula}) and (\ref{eq:abx}) for computing $\mS$.
Thus, for the differentiation of $\mS$, it seems also needed to compute
the derivatives of the eigenvalues $\bm{\Gamma}$ and eigenvectors $\mW$.

Unfortunately, the derivatives of eigenvectors in many cases are ill-defined.
Most notable is when there exist \textit{repeated} eigenvalues.  
Repeated eigenvalues are not uncommon: many photonic devices have certain structural
symmetries, from which eigenvalue repetition naturally emerges (see \figref{repeated_ev}).
For those repeated eigenvalues, their eigenvectors (up to a scale) are not
uniquely determined; any set of linearly independent vectors that span the
same subspace are valid eigenvectors. Because of the ambiguity, as the
structural parameter changes, those eigenvectors may change discontinuously
(see an examples in {Supplement 1}), and thus their derivatives may not be well-defined.

As a result, one must carefully choose eigenvectors in the subspace of repeated
eigenvalues such that the eigenvectors change continuously with respect to the
structural parameter.  
This choice, however, is computationally expensive. 
As derived in~\cite{van2007computation},
to ensure well-defined eigenvector derivatives,
one must compute higher-order derivatives of the eigenvalues and 
the matrix $\mP\mQ$: if the repeated eigenvalues have repeated
derivatives up to the $n$-th order (see \figref{repeated_ev}), then 
derivatives up to the $(n+1)$-th
order of both eigenvalues and the matrix $\mP\mQ$ must be computed to
determine first-order eigenvector derivatives. 

\subsection{Differentiation without Resort to Eigenvector Derivatives}\label{sec:der_alg}
We now present 
a new algorithm for computing the scattering matrix derivative $\mS'$.
Even in the presence of repeated eigenvalues and their derivatives, 
our method requires only the first-order derivatives of the matrices $\mP$ and $\mQ$,
completely sidestepping the differentiation of eigenvalues and eigenvectors.
In comparison to the way that takes eigenvalue derivatives (as described above),
our method 
is more robust and efficient.

First, we rewrite the commonly used expressions of scattering matrix
components, shown in~\eq{s_formula}, in new forms,
\begin{subequations}\label{eq:s_formula2}
  \begin{align}
    \mR_{\textrm{L}} = \mR_{\textrm{R}} & = \left(\mI -\mD_1^2\right)^{-1}\left(\mD_1\mD_2 - \mD_3\right),\label{eq:s11_2} \\
    \mT_{\text{LR}}=\mT_{\text{RL}} & = \left(\mI -\mD_1^2\right)^{-1}\left(\mD_2 - \mD_1\mD_3\right), 
  \end{align}
\end{subequations}
where $\mD_1$, $\mD_2$, and $\mD_3$ denote the following matrix
multiplications, respectively: 
\begin{subequations}\label{eq:D_mat}
\begin{align}
  \mD_1 & \coloneqq\mA^{-1}\mX\mB = \left(\mW_0+\mT\mV_0\right)^{-1}\mW\mX\mW^{-1}\left(\mW_0-\mT\mV_0\right),\\
  \mD_2 & \coloneqq\mA^{-1}\mX\mA = \left(\mW_0+\mT\mV_0\right)^{-1}\mW\mX\mW^{-1}\left(\mW_0+\mT\mV_0\right),\\
  \mD_3 & \coloneqq\mA^{-1}\mB = \left(\mW_0+\mT\mV_0\right)^{-1}\left(\mW_0-\mT\mV_0\right).
\end{align}
\end{subequations}
The derivation of these new expressions~(\ref{eq:s_formula2}) and~(\ref{eq:D_mat})
are provided in {Supplement 1}.
In Eqs.~(\ref{eq:D_mat}), the equalities are reached by applying~\eq{abx}
and using $\mT$ that denotes $\mT\coloneqq\bm{\Omega}\mQ^{-1}$.

The expressions in~\eq{s_formula2} present a new route for computing
scattering matrix derivative.
They indicate that the scattering matrix $\mS$ is determined by 
the three matrices, $\mD_1$, $\mD_2$, and $\mD_3$. 
As a result, to compute its derivative $\mS'$ using the chain rule,
we need to compute the derivatives of $\mD_1$, $\mD_2$, and $\mD_3$ 
with respect to the structural parameter. 

In~Eqs.~(\ref{eq:D_mat}), both $\mW_0$ and $\mV_0$ (introduced in~\eq{abx}) are constant matrices.
Thus, the derivatives of $\mD_1$, $\mD_2$, and $\mD_3$ 
only depend on the derivatives of two other matrices in~Eqs.~(\ref{eq:D_mat}),
namely $\mT$ and $\mW\mX\mW^{-1}$.
We now describe how to compute the derivatives of the two matrices, respectively.

\paragraph{Derivative of $\mT$.}
The matrix $\mT\coloneqq\bm{\Omega}\mQ^{-1}$ is related to $\mQ$ and $\bm{\Omega}\coloneqq\left(\mP\mQ\right)^{{1}/{2}}$
but not the eigenvalues and eigenvectors.
Its derivative can be expressed as
\begin{equation}
    \mT' = \bm{\Omega}'\mQ^{-1}+\bm{\Omega}\left(\mQ^{-1}\right)' = \bm{\Omega}'\mQ^{-1} - \bm{\Omega}\mQ^{-1}\mQ'\mQ^{-1},
\end{equation}
where $\mQ$ depends on the material permittivity
distributions of the photonic structure, and therefore its
derivative $\mQ'$ with respect to a design parameter
can be directly computed (see examples in \secref{ret}).
The way of computing $\bm{\Omega}'$ can be derived by taking the derivatives 
on both sides of the relation $\bm{\Omega}^2=\mP\mQ$, which yields
\begin{equation}\label{eq:omega_prime}
    \bm{\Omega}'\bm{\Omega} + \bm{\Omega}\bm{\Omega}' = \mP'\mQ + \mP\mQ'.
\end{equation}
Given $\mP'$ and $\mQ'$, the right-hand side of this equation can be directly computed. 
To compute $\bm{\Omega}'$, we rewrite the left-hand side by denoting
$\bm{\Omega}'$ as $\bm{\Omega}'=\mW\mY\mW^{-1}$ for some unknown $\mY$,
where $\mW$ is the eigenvector matrix of $\mP\mQ$. Using the fact that
$\bm{\Omega} = (\mP\mQ)^{1/2}=\mW\mLam\mW^{-1}$, we obtain a simplified form 
of~\eq{omega_prime}:
\begin{equation}\label{eq:Y}
    \mY\mLam + \mLam\mY = \mW^{-1}(\mP'\mQ+\mP\mQ')\mW.
\end{equation}
From~\eq{Y}, $\mY$ can be easily solved by noticing that $\mLam$ is a diagonal
matrix, and therefore \eq{Y} can be written element-wise as $(\lambda_i +
\lambda_j)\mY_{ij}=\mC_{ij}$, where $\lambda_i$ is the $i$-th eigenvalue in $\mLam$,
and $\mC$ denote the matrix on the right-hand side of~\eq{Y}.
In other words, the elements of $\mY$ can be obtained by solving $n^2$ 1D linear equations
in parallel. Once $\mY$ is obtained, $\bm{\Omega}'$ is computed using
$\bm{\Omega}'=\mW\mY\mW^{-1}$.

We note that while this process of computing $\bm{\Omega}'$ requires the
eigenvectors $\mW$ and eigenvalues $\mLam$, they are also needed for computing
the scattering matrix in the first place. 
Our solving process does not require the derivatives of eigenvectors.
Therefore, it introduces no additional effort in terms of eigen-decomposition. 

\paragraph{Derivative of $\mW\mX\mW^{-1}$.}
In the first glance, the derivative of $\mW\mX\mW^{-1}$
depends on the eigenvectors $\mW$.
However, from the definition of $\mX$ in~\eq{abx-a}, we notice that 
$\mW\mX\mW^{-1}=e^{j\bm{\Omega}{L}/{k_0}}$,
which suggests an alternative approach: take the derivative 
of the matrix exponential $e^{j\bm{\Omega}{L}/{k_0}}$ with respect to $\bm{\Omega}$.

A common approach of computing the matrix exponential 
$e^{j\bm{\Omega}{L}/{k_0}}$ is through the eigen-decomposition
of $\bm{\Omega}$ followed by the exponential of the resulting eigenvalues. 
If we take this approach, the derivative computation must involve 
the derivatives of eigenvectors, which might not be well-defined. 
Another approach, used by Feynman~\cite{feynman1951operator}
and others~\cite{karplus1948note,bellman1997introduction,snider1964perturbation}, expresses the derivative of a matrix exponential using
an integral that in itself involves matrix exponentials.
Yet, numerically evaluating the matrix exponentials and the integral are expensive.

Instead, our proposed method for computing the derivative is based on the following
proposition originally proved in~\cite{najfeld1995derivatives}. 
\begin{proposition}
Consider an $n\times n$ matrix $\bm{\Omega}$ and its derivative $\bm{\Omega}'$ with respect to an arbitrary parameter.
If 
\begin{equation}
\mG=\begin{bmatrix}\bm{\Omega} & \bm{\Omega}' \\ 
    \mathbf{0} & \bm{\Omega} 
\end{bmatrix},\;\textrm{then }
e^{j\mG {L}/{k_0}} =
\begin{bmatrix}
    e^{j\bm{\Omega}{L}/{k_0}} & \left(e^{j\bm{\Omega}{L}/{k_0}}\right)' \\
    \mathbf{0} & e^{j\bm{\Omega}{L}/{k_0}} 
\end{bmatrix},
\end{equation}
where the top-right $n\times n$ block matrix in $e^{j\mG {L}/{k_0}}$ is the derivative 
of the matrix exponential $e^{j\bm{\Omega}{L}/{k_0}}$.
\end{proposition}

In our problem, $\bm{\Omega}'$ is computed as described above (by solving~\eq{Y}),
and the common way of computing $e^{j\mG {L}/{k_0}}$ is by taking the 
eigen-decomposition of $\mG$, which is again what we wish to avoid.
We therefore take a different approach, the scaling and squaring method~\cite{higham2005scaling},
to compute $e^{j\mG {L}/{k_0}}$---without the need of eigen-decomposition.

The scaling and squaring method exploits the relation
$e^{\mA} = \left(e^{\mA/\sigma}\right)^\sigma$ for any $n\times n$ matrix $\mA$.
In practice, $\sigma$ is chosen to be $\sigma=2^s$ for some non-negative integer $s$.
The idea is to have the norm of $\mA/\sigma$ sufficiently small such that
$e^{\mA/\sigma}$ can be well approximated by a Pad\'{e} approximant near the
origin. The Pad\'{e} approximant is a rational polynomial of $\mA$. 
Its evaluation requires only matrix multiplications and inverse, but no eigen-decomposition.
The scaling and squaring method is robust and accurate, 
and has been used in many numerical tools (such as MATLAB's \textsf{expm} function). 

When applying this method, 
we further exploit the specific structure of $\mG$ (i.e., its bottom-left block
matrix vanishes, and its two diagonal block matrices are identical) to tailor
the method for improving computational performance.
{Supplement 1} presents our detailed derivations and computational steps. 

\begin{figure}[t]
\vspace{-3mm}
\centering
\includegraphics[width=\columnwidth]{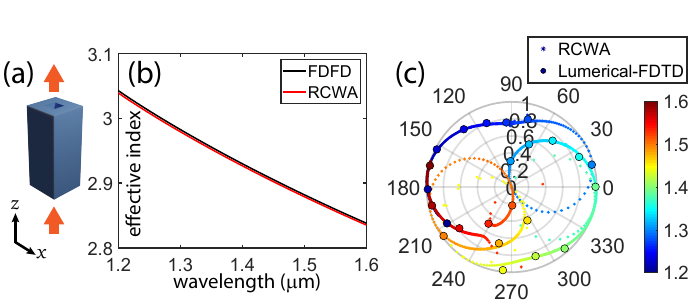}%
\vspace{-2.5mm}
\caption{\textbf{Accuracy comparison.}
    \textbf{(a)} We use our method and FDTD (Lumerical~\cite{solutions2003lumerical}) to analyze a 3D meta-atom. 
The width of the pillar and its square hole are $0.6 \mu\text{m}$ and $0.2\mu\text{m}$, respectively, 
and it has a height of 1.4 $\mu\text{m}$.
We use the periodic boundary condition, with the period of $0.66\mu\text{m}$. 
\textbf{(b)} We scan the (x-polarized) wavelength and plot the effective indices of the fundamental mode evaluated by
FDFD and our method. 
\textbf{(c)} For each wavelength (color mapped here),
we compute the far-field amplitudes and phases changes, and compare our 
results to FDTD. 
\label{fig:preliminary}}
\vspace{-2mm}
\end{figure}

\section{Results}\label{sec:ret}
This section presents our numerical results.
First, we validate our algorithm for computing scattering matrix derivatives.
Next, to demonstrate the use of scattering matrix derivatives in photonics,
we optimize the geometry of photonic metasurface units (also called
\textit{meta-atoms}).

Meta-atoms are the building blocks of a metasurface,
often designed based on physical intuitions and manually crafted
libraries~\cite{yu2012broadband, overvig2019dielectric, overvig2018two}.
More recently, inverse design methods of meta-atom structures have also been 
explored---e.g., through finite-difference-based gradient descent~\cite{backer2019computational},
adjoint-based level-set method~\cite{mansouree2019metasurface}, and
topological optimization~\cite{sell2017large,lin2019topology}.

Due to fabrication constraints, meta-atoms often have constant cross-sectional
shapes along one direction (i.e., $z$-direction, as shown in \figref{preliminary}-a).
Thus, the semi-analytical methods (such as RCWA) are particularly efficient for
simulating meta-atoms, thanks to their ability of \textit{not} discretizing along
$z$-direction~\cite{moharam1981rigorous}. Our method, for the first time, enables the semi-analytical methods
to also compute scattering matrix derivatives with respect to design parameters.
Here, in the framework of RCWA, we demonstrate 
automatic discovery of meta-atom structures that reach various amplitude and phase goals.

\subsection{Validation}
To validate our algorithm, we consider a {dielectric} meta-atom used in
metasurface holography~\cite{overvig2018two, overvig2019dielectric}. 
Its structure is shown in \figref{preliminary}-a. 
We use Eqs.~(\ref{eq:s_formula2}) to compute the 
scattering matrix, for which the matrices $\mP$ and $\mQ$ (introduced in~\eq{pq_form})
are constructed using RCWA. The scattering matrix allows us to compute
light propagation properties of the meta-atom, which are in turn compared 
to the results from finite difference time domain (FDTD) method implemented
in \textsf{Lumerical}~\cite{solutions2003lumerical}.

We first scan the light wavelength from $1.2 \mu\text{m}$ to $1.6 \mu\text{m}$. For
each wavelength, we compute, using our scattering matrix and FDFD respectively, 
the effective index of the fundamental mode propagating in the meta-atom.
The results from our method agree with FDFD results (see \figref{preliminary}-b). 
Furthermore, we consider the far-field light transmission through the meta-atom,
and compute the phase shift and amplitude change for each wavelength. Again, the
results from our method and FDTD match closely, as shown in \figref{preliminary}-c.

These experiments confirm that our scattering matrix computation is as accurate as 
FDTD in \textsf{Lumerical}. In terms of computational cost, our 
method takes about 0.15 seconds for each monochromatic simulation, and a few seconds
for the entire $1.2 \mu\text{m}$-$1.6 \mu\text{m}$ wavelength range, whereas the FDTD simulation takes several minutes.


Next, we validate our derivative computation. We
consider again the meta-atom structure shown in \figref{preliminary}-a, and choose 
the parameter $\alpha$ to be the size of the hollow square.
Using our method, we compute the derivative of the structure's scattering matrix
with respect to $\alpha$. Meanwhile, since there is no analytic expression of the scattering matrix derivative,
we approximate it using finite difference (FD) estimation, that is,
 \begin{equation}
 \frac{\partial \mS}{\partial \alpha} \approx \frac{\mS(\alpha + \Delta\alpha) - \mS(\alpha - \Delta\alpha)}{2\Delta\alpha}.
 \end{equation}
We estimate $\frac{\partial \mS}{\partial \alpha}$ using a sweeping range of $\Delta \alpha$ values, and compare them
to the derivative resulted from our method.

\begin{figure}[t]
\centering
\includegraphics[width=0.9\columnwidth]{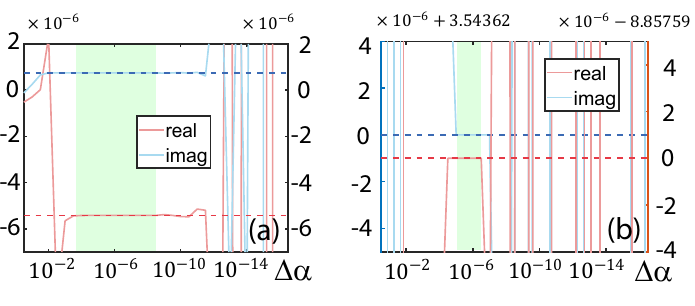}%
\vspace{-4mm}
\caption{\textbf{Scattering matrix derivatives.}
We choose two matrix elements in the scattering matrix, and plot
their FD derivatives estimated with different FD sizes ($\Delta\alpha$ in $x$-axis)
in (a) and (b), respectively. 
In each plot, the red and blue solid curves correspond to the real and imaginary parts of
the estimated derivative. 
Meanwhile, the derivatives computed by our method are 
indicated by the red (real) and blue (imaginary) horizontal dash lines.
These plots show that FD estimation is highly sensitive to $\Delta\alpha$. 
Light green regions indicate valid $\Delta\alpha$ ranges for both matrix elements.
The valid $\Delta\alpha$ varies element by element.
Thus, in practice, it is hard to choose a proper $\Delta\alpha$ for the entire scattering matrix, whereas our method is always robust.
\label{fig:test_fd}}
\vspace{-2mm}
\end{figure}

\begin{figure}[t]
\vspace{-4mm}
\centering
\includegraphics[width=0.8\textwidth]{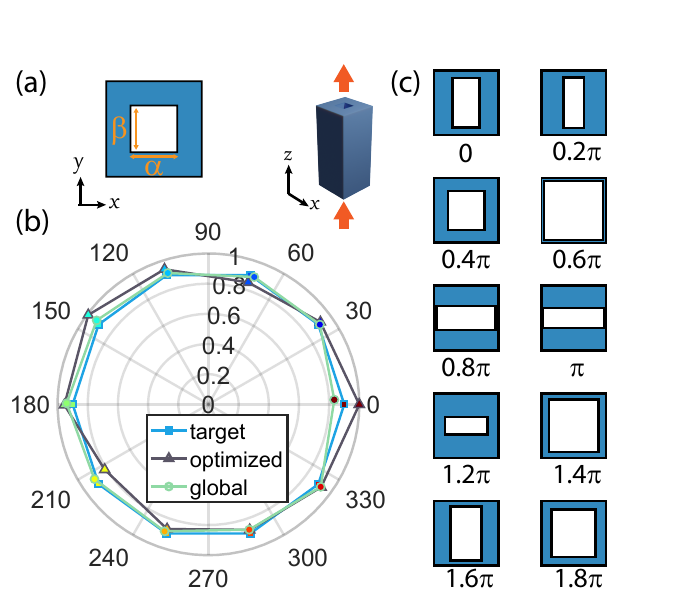}%
\vspace{-4mm}
\caption{\textbf{Optimization of \eq{ex1}.} 
(a) We optimize the cross-sectional shape specified by two design parameters $\alpha$ and $\beta$
in order to reach a target transmission amplitude and phase.
We examine different targets evenly sampled on a circle on the complex plane
(indicated by the square dots in (b)). The optimized amplitudes and phases (indicated by triangular dots) reach 
closely to the targets. As a reference, we also show the amplitudes and phases (in circular dots) 
of the designs that globally minimize~\eq{ex1}, that is, ones obtained
through a slow, exhaustive search of all parameter combinations.
While no gradient-based optimization algorithm can guarantee the global minimum 
of~\eq{ex1}, our results approach the targets closely, comparable to what the global 
minimums can achieve. The resulting cross-sectional shape for each sampled target are shown in (c).
\label{fig:pa_results}}
\vspace{-2mm}
\end{figure}

The results are illustrated in \figref{test_fd}. The accuracy of FD approximation 
largely depends on the choice of $\Delta\alpha$.
Only when $\Delta\alpha$ is chosen within a certain range,
FD approximation is accurate enough to agree with our derivative results.
This agreement confirms the correctness of our method.
But for different elements in the scattering matrix, the valid $\Delta\alpha$ range
varies (indicated in light green in \figref{test_fd}), suggesting that FD approximation is impractical:
it is hard, if not impossible, to choose a proper $\Delta\alpha$ to produce accurate
derivative estimations for all elements in the scattering matrix.
In contrast, our method is robust for computing the derivatives.

\begin{figure*}[t]
\centering
\includegraphics[width=0.83\textwidth]{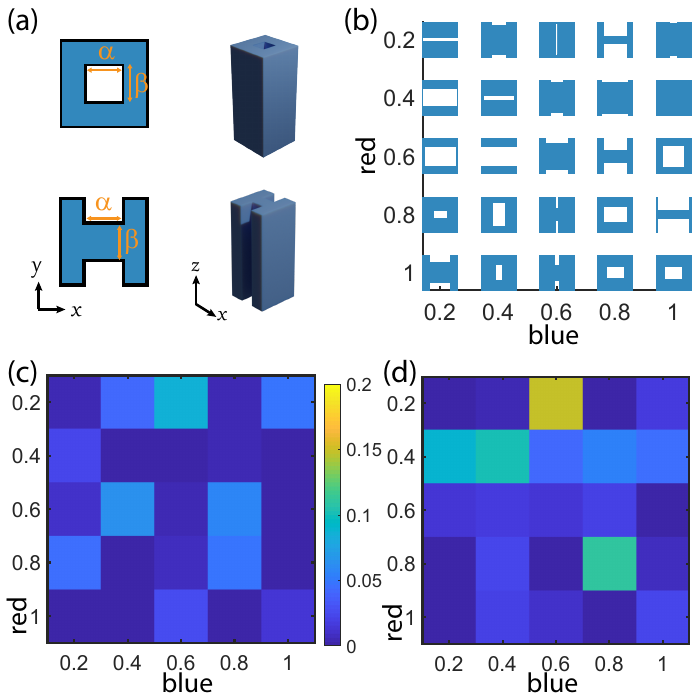}%
\vspace{-3mm}
\caption{\textbf{Optimization of \eq{ex3}.}
(a) We optimize meta-atom structures described by two archetypes,
each with two parameters.
The goal is to obtain desired amplitude responses at two separate wavelengths (i.e.,
1.2$\mu\text{m}$ (\textsf{blue}) and 1.6$\mu\text{m}$ (\textsf{red})), simultaneously.
We sample five amplitudes from 0.2 to 1 for each wavelength, forming 25 different
optimization targets. Each target leads to a different cross-sectional design shown in (b).
For \textsf{red} light, the discrepancies between achieved amplitudes and the targets
are shown in (c), and the same visualization for \textsf{blue} light is shown in (d).
\label{fig:amp2d}}
\vspace{-5mm}
\end{figure*}

\paragraph{Computational cost.}
In addition to the robustness, our method is also faster than the FD method.
In the FD method, computing a matrix derivative requires the computation of two scattering 
matrices $\mS(\alpha+\Delta\alpha)$ and $\mS(\alpha-\Delta\alpha)$. In contrast,
our method, in addition to computing $\mS(\alpha)$,
only requires a few matrix multiplications and inverses (recall Section 3.\ref{sec:der_alg}). 
On our workstation computer, the overhead of computing a scattering matrix derivative
is about $30\%\sim40\%$ of the cost for computing the scattering matrix itself.

\begin{figure*}
\centering
\hspace{10mm}
\includegraphics[width=0.8\textwidth]{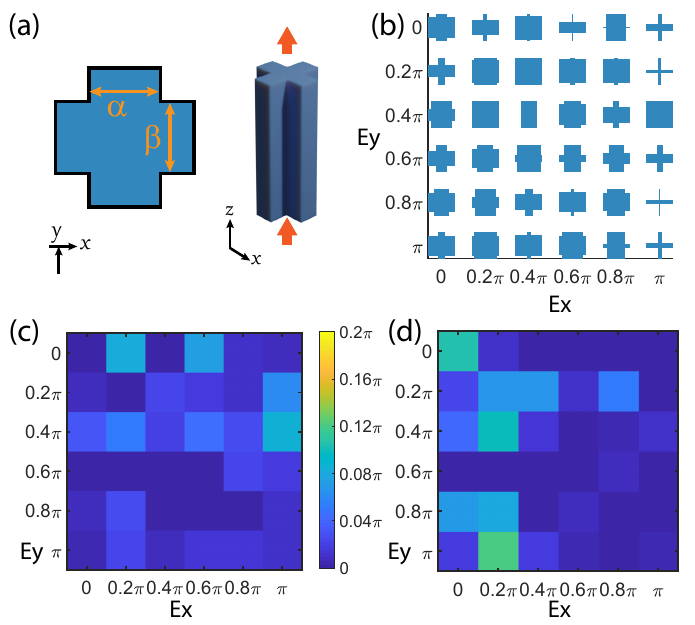}%
\vspace{-4mm}
\caption{\textbf{Optimization of~\eq{ex2}.} 
(a) We optimize the meta-atom structure described by two parameters. 
The goal is to achieve certain phase changes for $x$- and $y$-polarized light
simultaneously. (b) We sample six target phase changes for $x$- and
$y$-polarized light, respectively. Their combination forms 36 different optimization targets.
For each target, our optimization produces a cross-sectional shape design.
(c) For $x$-polarized incident light, we show the residual (in terms of phase
angle difference) between each pair of inversely designed phase change and the target.
And similar visualization for $y$-polarized light is shown in (d). 
\label{fig:dual_phase}}
\vspace{-1mm}
\end{figure*}

\subsection{Use Case: Optimization of Meta-atom Structure}
\paragraph{Controlling phase and amplitude of monochromatic light.}
First, we optimize meta-atom structures to reach specific transmitted amplitudes and phases
for a monochromatic light (at 1.55$\mu\text{m}$ wavelength, $x$-polarized).
The cross-sectional shape is shown in \figref{pa_results}-a, determined by two parameters.
The objective function for the inverse design is defined as
\begin{equation}\label{eq:ex1}
    \mathcal{L} = {\left| \mT_{\text{LR}}(m, m) - t_m\right|}^2,
\end{equation}
where $\mT_{\text{LR}}$ is the transmission submatrix in the scattering matrix (recall~\eq{scattering}),
$m$ is the mode index for the incident and outgoing light in free space, 
thus $\mT_{\text{LR}}(m, m)$ denotes the $m$-th diagonal element of the matrix.
Also, $t_m$ is a complex constant specifying the target amplitude and phase of
the transmission.  
Here we consider the fundamental mode (the way to choose corresponding $m$ is given in Supplemental 1), 
which describes the far-field light transmission along the $z$-direction.

To verify the robustness of our method and the enabled optimization,
we evenly sample different targets $t_m$ on a circle {on} the complex plane 
(see \figref{pa_results}-b).  For each target, we find meta-atom's shape parameters by minimizing~\eq{ex1}
through a gradient-descent algorithm~\cite{ruder2016overview},
for which the gradients of \eq{ex1} with respect to the design parameters are
computed using our method.
As shown in \figref{pa_results}-b, we are able to automatically discover structures
that reach these targets closely.

\paragraph{Controlling phases for both $x-$ and $y-$polarized light.}
Next, we optimize meta-atom structures to obtain target responses for
$x$- and $y$-polarized light, simultaneously. This type of meta-atoms 
has been used to construct metasurface holograms~\cite{chen2014high}.
In our example, the light wavelength is 1.3$\mu\text{m}$; the meta-atoms have
a fixed height of 2.0$\mu\text{m}$ and a period of 2.5$\mu\text{m}$ along $x$- and $y$-direction.
The cross-sectional shape of the meta-atoms are specified by two parameters shown in 
\figref{dual_phase}-a. 
We determine the parameters by minimizing the following objective function:
\begin{equation}\label{eq:ex2}
    \mathcal{L} = -\frac{\mT_{\text{LR}}(m_x, m_x)}{\left|\mT_{\text{LR}}(m_x, m_x)\right|}t_x^*-\frac{\mT_{\text{LR}}(m_y, m_y)}{\left|\mT_{\text{LR}}(m_y, m_y)\right|}t_y^*,
\end{equation}
where the subscript $x$ (and $y$) indicates light polarization; 
$t_x$ (and $t_y$) are the target phase changes from $x$-polarized (and $y$-polarized) incident light to 
the outgoing light with the same polarization (i.e., $t_x=\exp(i\phi_x)$ for some $\phi_x$).
The first term in \eq{ex2} measures, for the $x$-polarized light, the cosine difference (through dot product on complex plane)
between the $m$-th mode's phase change and the target phase change,
and similarly for the second term.
The optimized structures for different $x$- and $y$-polarized phase targets are 
shown in \figref{dual_phase}. In all cases, the residual between the target 
and the resulting phase change is within 7\% of one period ($2\pi$), and in most cases within 1\%.

\paragraph{Controlling amplitudes for multiple wavelengths.}
We also demonstrate inverse design of meta-atoms for another type of optical response:
obtain two target amplitude responses at two separate wavelengths, simultaneously.
This type of responses have proven useful for making colored metasurface 
holograms~\cite{overvig2018two,overvig2019dielectric}.
Here we consider two archetypes used in~\cite{overvig2019dielectric}, each described 
by two parameters (see \figref{amp2d}-a). The two wavelengths under consideration are
1.2$\mu\text{m}$ (labeled as \textsf{blue}) and 1.6$\mu\text{m}$ (\textsf{red}), and the objective function is defined as
\begin{equation}\label{eq:ex3}
    \mathcal{L} = \left[ \left|\mT_{\text{LR},1}(m, m)\right|^2 - A^2_{1}\right]^2
    + \left[ \left|\mT_{\text{LR},2}(m, m)\right|^2 - A^2_{2}\right]^2.
\end{equation}
Here the subscript ``1'' and ``2'' indicate the \textsf{blue} (1.2$\mu\text{m}$) and \textsf{red} (1.6$\mu\text{m}$) wavelength, respectively.
The first term accounts for the \textsf{blue} wavelength:
$\mT_{\text{LR},1}$ is the transmission submatrix of the scattering matrix 
and $A_1$ is the desired amplitude. Similar is the second term.
More terms can be added in~\eq{ex3} to incorporate more than two wavelengths.

For each archetype, we find its parameter values via a {gradient-descent} algorithm that minimizes 
\eq{ex3}, and choose between the two archetypes one that produces a smaller objective value.
The optimized structures and their performances are shown in \figref{amp2d}.
For almost all the experiments (each with a different amplitude target),
the resulting amplitudes by the inversely designed meta-atoms match closely to
their targets. 

\begin{figure}[t]
\centering
\includegraphics[width=0.8\textwidth]{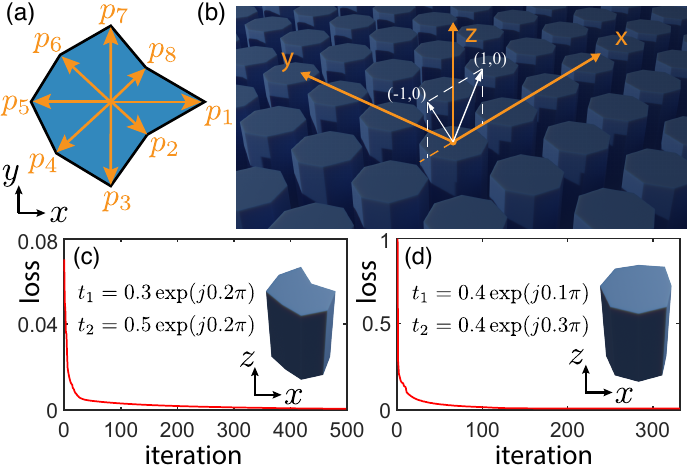}%
\vspace{-2mm}
\caption{\textbf{Inverse design of star-convex meta-atoms.}
(a) The star-convex polygon is used to represent the cross-section of a meta-atom,
defined by many control variables ($p_1$\ldots$p_8$ in this case).
As an example, we inverse design the shape for reaching target amplitudes and phases
in two scattering directions (i.e., corresponding to diffraction orders (-1,0) and (1,0) in (b)) simultaneously.
(c-d) We perform two experiments to reach two sets of $(t_1, t_2)$ goals shown in the plots.
In each experiment, our optimization finds the design parameters within
hundreds of iterations, resulting
in nontrivial shapes that are hard to be manually designed.
\label{fig:asym_opt}}
\end{figure}

\paragraph{General cross-sectional shape design.}
Lastly, we introduce a new way to inverse design the meta-atom's
cross-sectional shape under a general representation.
We use the star-convex polygon~\cite{stanek1977characterization} to represent the cross-sectional shape. 
Such a shape can be discretized by sampling $N$
points on its boundary so that the polar angles of these points are evenly
distributed over $[0, 2\pi]$. In other words, 
the $(k+1)$-th point has the coordinate $p_k\left[\cos{\left( 2k\pi/N\right)},
-\sin{\left( 2k\pi/N\right)}\right]$, where $p_k$ is a non-negative value (see
\figref{asym_opt}-a), and the shape is specified by $N$ parameters $p_1,\ldots,p_N$.
A large $N$ offers many degrees of freedom to represent a complex shape,
but meanwhile renders exhaustive search through the entire parameter space 
too expensive---one must rely on numerical optimization methods to determine the parameter 
values. 

This shape representation is particularly suitable for RCWA-based analysis, as it allows for a closed-form
2D Fourier transform of the shape (and thus the permittivity distribution)~\cite{lee1983fourier}.
In RCWA framework, 2D Fourier transform of the cross-sectional permittivity distribution
is needed for computing the matrices, $\mP$ and $\mQ$, as well as their derivatives with respect
to the $p_k$ parameters. Supplemental 1 provides the details of this process.

As examples, we optimize octagons ($N=8$) to obtain desired optical responses
in different scattering directions. 
First, we specify the target scattering directions.
Notice that to predict optical behavior of a single meta-atom in simulation,
periodic boundary condition is often used. Under this condition, the meta-atom
is effectively a 2D grating structure, for which 
we can use diffraction orders to specify different scattering directions:
the output light with diffraction order $(p, q)$ is along the direction
\begin{equation}
\vec{k} = \left(\frac{2\pi p}{L_x}, \frac{2\pi q}{L_y}, 1 \right),
\end{equation}
where $L_x$ and $L_y$ are periods along $x$- and $y$-axis, respectively ($L_x=L_y=1\mu\text{m}$ in our examples). 

We consider $x$-polarized light with the wavelength of 1.55\um.
The goal here is to obtain specified far-field phases and amplitudes
at two scattering directions---ones that correspond to the diffraction orders, $(-1,0)$
and $(1, 0)$, as shown in \figref{asym_opt}-b.
We further restrict $p_k$ to be in the range $[0.15\mum,0.45\mum]$, and determine
$p_k$ values by minimizing 
\begin{equation}
    \mathcal{L} = {\left| \mT_{\text{LR}}(m, n_1) - t_1 \right|}^2 + {\left| \mT_{\text{LR}}(m, n_2) - t_2 \right|}^2,
\end{equation}
where $n_1$ and $n_2$ are mode indices for the diffraction orders $(-1,0)$ and $(1,0)$, respectively;
and $t_1$ and $t_2$ specify the target phases and amplitudes (as complex values) in the two outgoing directions.
We perform two experiments for two sets of $t_1$ and $t_2$ goals.
The optimization convergence curves and resulting shapes are shown in \figref{asym_opt}.

\vspace{-0.5mm}
\section{Conclusion}
\vspace{-0.5mm}
We have presented an algorithm for computing the derivatives of the scattering 
matrices of a photonic structure with respect to its structural parameters.
Our method is built on the framework of semi-analytical methods for analyzing photonic
structures. A key step in semi-analytical methods
for computing scattering matrices is the eigen-decomposition. 
However, 
to compute scattering matrix derivatives, 
directly differentiating the eigenvalue analysis 
poses significant difficulties.
We show a new route to compute scattering matrix derivatives
without the need of differentiating the eigen-decomposition process.

The scattering matrix derivatives present how a photonic structure's
performance changes as its structural parameters vary. 
While we demonstrated their use in the optimization of meta-atom units,
they are useful in many other applications.
Therefore, our method may serve as a useful analysis tool in a wide range of 
photonic design tasks.

\vspace{-1mm}
\section*{Funding}
National Science Foundation (CAREER-1453101, 1717178, 1816041).

\vspace{-2mm}
\section*{Acknowledgments}
We thank Nanfang Yu for valuable suggestions.

\vspace{-2mm}
\section*{Disclosures}
The authors declare no conflicts of interest.

\vspace{1mm}
\noindent
See Supplement 1 for supporting content.

\bibliographystyle{acm}
\bibliography{ref}



\section*{Supplementary material}

# Differentiable Scattering Matrix for Optimization of Photonic Structures: supplemental document

Ziwei Zhu[1] and Changxi Zheng[1,*]

[1]Department of Computer Science, Columbia University, New York, New York 10027, USA
*Corresponding author: cxz@cs.columbia.edu

September 23, 2020

## Abstract

This document provides supplementary information to the paper "Differentiable Scattering Matrix for Optimization of Photonic Structures".

# 1 Matrix Construction in RCWA

## 1.1 Construct Wave Equation Coefficient Matrices

When analyzing a meta-atom structure, we use $z$-direction to indicate the wave propagation direction, and use periodic boundary condition on $x$-$y$ plane. In the framework of RCWA, the electric and magnetic fields at a distance $z$ are represented under a set of Fourier basis functions (also called harmonics).

Suppose the E- or H-field is represented using $n_x \times n_y$ harmonics. The $i$-th harmonic along the $x$-axis together with the $j$-th harmonic along the $y$-axis describes a wave vector projected on the $x$-$y$ plane, that is,

$$\vec{k}_{i,j} = (k_{ix}, k_{jy})$$
$$= \left(\frac{2\pi}{L_x}\left(i - \frac{n_x+1}{2}\right), \frac{2\pi}{L_y}\left(j - \frac{n_y+1}{2}\right)\right) \quad (1)$$

where $L_x$ and $L_y$ are periods along $x$- and $y$-axis, respectively. When $i = \frac{n_x+1}{2}$ and $j = \frac{n_y+1}{2}$, the wave vector $\vec{k}_{i,j} = \vec{0}$, corresponding to the wave propagating only along $z$-direction.

With these Fourier basis functions, the E-field at a distance $z$ can be expressed as

$$E(x, y, z) = \sum_{i=1}^{n_x}\sum_{j=1}^{n_y} c_{i,j}(z) e^{j(k_{ix}x + k_{jy}y)}, \quad (2)$$

where $c_{i,j}(z)$ are Fourier coefficients, and they are functions of $z$. In other words, (2) is a representation using separation of variables. It discretizes the field on $x$-$y$ plane but remain analytic along $z$-direction. The H-field can be (semi-)discretized in the same way.

Substituting (2) into the frequency-domain Maxwell's equations leads to the semi-discussed equations (2) in the main text. Here, for the sake of simplicity of the presentation and following the use in common situations, we assume the materials are non-ferromagnetic. Then, the coefficient matrices $\mathbf{P}$ and $\mathbf{Q}$ in Eqn. (2) have the following forms,

$$\mathbf{P} = \begin{bmatrix} \mathbf{K}_x[[\epsilon]]^{-1}\mathbf{K}_y & k_0^2\mathbf{I} - \mathbf{K}_x[[\epsilon]]^{-1}\mathbf{K}_x \\ \mathbf{K}_y[[\epsilon]]^{-1}\mathbf{K}_y - k_0^2\mathbf{I} & -\mathbf{K}_y[[\epsilon]]^{-1}\mathbf{K}_x \end{bmatrix} \text{ and}$$

$$\mathbf{Q} = \begin{bmatrix} -\mathbf{K}_x\mathbf{K}_y & \mathbf{K}_x\mathbf{K}_x - k_0^2[[\epsilon]] \\ k_0^2[[\epsilon]] - \mathbf{K}_y\mathbf{K}_y & \mathbf{K}_y\mathbf{K}_x \end{bmatrix},$$

(3)

where $k_0 = 2\frac{\pi}{\lambda}$ for the considered wavelength $\lambda$, $[[\epsilon]]$ is the matrix describing the permittivity distribution on $x$-$y$ plane; its construction will be presented in the next subsection. Lastly, $\mathbf{K}_x$ and $\mathbf{K}_y$ are both diagonal matrices of size $n_xn_y \times n_xn_y$ [3], and the $[j - 1 + (i - 1)n_y]$-th diagonal elements in $\mathbf{K}_x$ and $\mathbf{K}_y$ are expressed, respectively, as follows:

$$\mathbf{K}_x(j - 1 + (i - 1)n_y) = \frac{2}{L_x}\pi\left(i - \frac{n_x + 1}{2}\right) \text{ and}$$

$$\mathbf{K}_y(j - 1 + (i - 1)n_y) = \frac{2}{L_y}\pi\left(j - \frac{n_y + 1}{2}\right).$$

(4)

In the expressions of $\mathbf{P}$ and $\mathbf{Q}$, $k_0$, $\mathbf{K}_x$, and $\mathbf{K}_y$ are all independent from the structural design parameters; they are constant values when one take the derivatives with respect to design parameters. The permittivity matrix $[[\epsilon]]$, however, will change with respect to the design parameters. Therefore, the derivatives of the coefficient matrices have the forms,

$$d\mathbf{P} = \begin{bmatrix} \mathbf{K}_x d[[\epsilon]]^{-1}\mathbf{K}_y & -\mathbf{K}_x d[[\epsilon]]^{-1}\mathbf{K}_x \\ \mathbf{K}_y d[[\epsilon]]^{-1}\mathbf{K}_y & -\mathbf{K}_y d[[\epsilon]]^{-1}\mathbf{K}_x \end{bmatrix} \text{ and}$$

$$d\mathbf{Q} = \begin{bmatrix} \mathbf{0} & -k_0^2 d[[\epsilon]] \\ k_0^2 d[[\epsilon]] & \mathbf{0} \end{bmatrix}$$

(5)

where the symbol d indicates the derivative with respect to a particular design parameter $p$, and $d[[\epsilon]]^{-1}$ can be expressed as

$$d[[\epsilon]]^{-1} = -[[\epsilon]]^{-1}d[[\epsilon]][[\epsilon]]^{-1}.$$

(6)

## 1.2 Permittivity Matrix of Grid-discretized Shapes

To construct the permittivity matrix $[[\epsilon]]$ in (3), we consider the cross-sectional region of the photonic structure on $x$-$y$ plane. Suppose the region has a size

$L_x \times L_y$. The continuous Fourier transform of the permittivity distribution in this region is written as

$$\epsilon(m,n) = \frac{1}{S} \int \int_{\mathcal{S}} \epsilon(x,y) e^{\left(-i\frac{2\pi}{L_x}mx\right)} e^{\left(-i\frac{2\pi}{L_y}ny\right)} \mathrm{d}x \mathrm{d}y \qquad (7)$$

where $\mathcal{S}$ is the cross-sectional region and $S$ is its area.

To compute (7) numerically, we first divide the cross-sectional region into $P_x \times P_y$ grids. Assume that the permittivity in each grid $(l,p)$ is uniformly distributed, denoted as $\epsilon_{l,p}$. Then, the (discrete) Fourier transform of all grids is a summation of the Fourier transform of each grid, namely,

$$\epsilon(m,n) = \sum_{l=1}^{P_x} \sum_{p=1}^{P_y} \epsilon_{l,p} \left[ \frac{1}{L_x} \exp\left(-i\frac{m\pi}{L_x}(u_l + u_{l-1})\right) \operatorname{sinc}\left(\frac{m\pi}{L_x}(u_l - u_{l-1})\right) (u_l - u_{l-1}) \right]$$

$$\times \left[ \frac{1}{L_y} \exp\left(-i\frac{n\pi}{L_y}(v_p + v_{p-1})\right) \operatorname{sinc}\left(\frac{n\pi}{L_y}(v_p - v_{p-1})\right) (v_p - v_{p-1}) \right], \qquad (8)$$

where $\operatorname{sinc}(x) = \frac{\sin x}{x}$, $u_l$ is the $x$-coordinate of the $l$-th vertical grid line, and $v_p$ is the $y$-coordinate of the $p$-th horizontal grid line. $m$ and $n$ are indices of the discrete Fourier coefficients: $m = -(n_x - 1), \cdots, 0, 1, \cdots, (n_x - 1)$ and $n = -(n_y - 1), \cdots, 0, 1, \cdots, (n_y - 1)$.

After computing $\epsilon(m,n)$ for all $m$ and $n$, we can construct the permittivity matrix $[[\epsilon]]$, which has the size $n_x n_y \times n_x n_y$. The matrix $[[\epsilon]]$ has the so-called block-Toeplitz structure, containing matrix blocks that are repeated down the diagonals of the matrix. Its element values are assembled from the Fourier coefficients $\epsilon(m,n)$ in the following way:

$$[[\epsilon]](in_y + j, kn_y + l) = \epsilon(i-k, j-l), \qquad (9)$$

where the indices $i$ and $k$ are in the range $0, \ldots, n_x - 1$; $j$ and $l$ are in the range $0, \ldots, n_y - 1$.

Similarly, the derivative of $[[\epsilon]]$ with respect to a design parameter can be computed using

$$\mathrm{d}[[\epsilon]](in_y + j, kn_y + l) = \mathrm{d}\epsilon(i-k, j-l), \qquad (10)$$

where the scalar derivative $\mathrm{d}\epsilon(i-k, j-l)$ is computed by differentiating (8) with respect to the design parameter.

## 1.3 Permittivity Matrix of Star-Convex Shapes

For a star-convex shape discussed in Section 4.B (recall Fig. 8) of the main text, we do not need to discretize the shape into grid-based representation. Suppose the shape is described by the parameters $p_1, \ldots, p_N$ (recall Fig. 8-a in the main text). As derived in [2], in this case the Fourier transform (7) can be computed

directly:
$$\tilde{\epsilon}(m,n) = \frac{1}{S} \sum_{k=1}^{N} \underbrace{\exp\left(j\vec{w} \cdot \vec{p}_k\right) \frac{\hat{n} \times \vec{\alpha}_k \cdot \vec{\alpha}_{k-1}}{(\vec{w} \cdot \vec{\alpha}_k)(\vec{w} \cdot \vec{\alpha}_{k-1})}}_{E_k}, \tag{11}$$

where $\hat{n} = (0,0,1)$ is the $z$-axis vector, $\vec{w} = \left(-2\pi \frac{m}{L_x}, -2\pi \frac{n}{L_y}\right)$, and $\vec{\alpha}_k = \vec{p}_{k+1} - \vec{p}_k$.

There exist two corner cases in which the denominator in (11) vanishes, and thus they require special treatments:

1. When $m = n = 0$, the vector $\vec{w}$ in (11) vanishes. In this case, we resort to the continuous Fourier transform (7), which leads to $\tilde{\epsilon}(0,0) = 1$.

2. $\vec{w}$ may become perpendicular to $\vec{\alpha}_k$, and we have
$$\vec{w} \cdot \vec{\alpha}_k = 0 \text{ and } \vec{w} \cdot \vec{p}_{k+1} = \vec{w} \cdot \vec{p}_k. \tag{12}$$

In this case, we consider two terms $E_k$ and $E_{k+1}$ in the summation (11). It can be shown that $E_k + E_{k+1} = \frac{0}{0}$, which is mathematically undefined. Instead, $E_k + E_{k+1}$ should be defined as its limit as the parameter $p_k$ (i.e., the length of $\vec{p}_k$) approaches to the situation where $\vec{w} \cdot \vec{\alpha}_k = 0$ occurs. This limit can be computed using L'Hôpital's rule, by taking the derivatives of both the numerator and the denominator with respect to $p_k$. This process leads to the following form:

$$E_k + E_{k+1} = \exp\left(j\vec{w} \cdot \vec{p}_k\right) \left\{ -\frac{[\hat{p}_k \cdot (\hat{n} \times (\vec{a}_k + \vec{a}_{k-1}))]}{(\vec{w} \cdot \hat{p}_k)(\vec{w} \cdot \vec{a}_{k-1})} + \frac{[(\hat{n} \times \vec{a}_{k+1}) \cdot \hat{p}_k]}{(\vec{w} \cdot \hat{p}_k)(\vec{w} \cdot \vec{a}_{k+1})} \right.$$
$$\left. -\frac{j[(\hat{n} \times \vec{a}_k) \cdot \vec{a}_{k-1}]}{(\vec{w} \cdot \vec{a}_{k-1})} - \frac{[(\hat{n} \times \vec{a}_{k+1}) \cdot \vec{a}_k]}{(\vec{w} \cdot \vec{a}_{k-1})(\vec{w} \cdot \vec{a}_{k+1})} \right\} \tag{13}$$

Here it is safe to assume that $\vec{w}$ is *not* perpendicular to $\vec{\alpha}_{k-1}$ or $\vec{\alpha}_{k+1}$. This is because, if it is indeed perpendicular to $\vec{\alpha}_{k-1}$, then $\vec{p}_{k-1}$, $\vec{p}_k$, and $\vec{p}_{k+1}$ are co-linear, and we can discard the parameter $p_k$ while still representing the same star-convex shape. Similarly, if $\vec{w}$ is perpendicular to $\vec{\alpha}_{k+1}$, we can safely discard $p_{k+1}$.

Lastly, note that $\tilde{\epsilon}(m,n)$ in (11) is merely the Fourier transform of the star-convex shape, but not the entire material in an $L_x \times L_y$ region. The permittivity distribution in the entire region can be viewed as a superposition of a background permittivity $\epsilon_0$ and a star-convex shape with the permittivity $\epsilon_1 - \epsilon_0$. Thus the Fourier transform of the entire permittivity distribution is the Fourier transforms of both components:

$$\epsilon(m,n) = \begin{cases} \frac{\epsilon_1 - \epsilon_0}{L_x L_y} \tilde{\epsilon}(m,n) + \epsilon_0, \text{ if } m = n = 0; \\ \frac{\epsilon_1 - \epsilon_0}{L_x L_y} \tilde{\epsilon}(m,n), \text{ Otherwise.} \end{cases} \tag{14}$$

Once $\epsilon(m,n)$ for all $m$ and $n$ are obtained, we can assemble the permittivity matrix $[[\epsilon]]$ and compute its derivative $d[[\epsilon]]$ using the same formula (9) and (10) presented earlier.

## 2  An Example of Eigenvector Discontinuity

Figure 2 in the main text provides an exemplar meta-atom structure in which repeated eigenvalues (and thus repeated propagation constants) exist. Here we use a simple mathematical example to show that when repeated eigenvalues emerge, the eigenvector derivatives become undefined.

Consider the following $2 \times 2$ matrix with a parameter $p$:

$$\mathbf{A}(p) = \begin{bmatrix} 1 & p \\ p & 1 \end{bmatrix}. \tag{15}$$

When $p \neq 0$, its eigenvectors form a constant matrix.

$$\mathbf{X} = \begin{bmatrix} -\frac{\sqrt{2}}{2} & \frac{\sqrt{2}}{2} \\ \frac{\sqrt{2}}{2} & \frac{\sqrt{2}}{2} \end{bmatrix}. \tag{16}$$

In fact, when $p \neq 0$, this eigenvector matrix is unique up to a scale. When $p$ becomes zero, repeated eigenvalues emerge (both are 1). However, while $\mathbf{X}$ is still a valid eigenvector matrix of $\mathbf{A}(0)$, it is not unique anymore. In fact, any $2 \times 2$ orthonormal matrix is a valid eigenvector matrix of $\mathbf{A}(0)$. For example, the standard eigen-decomposition solver (e.g., in Matlab) gives

$$\mathbf{X}(0) = \begin{bmatrix} 1 & 0 \\ 0 & 1 \end{bmatrix}. \tag{17}$$

Therefore, in the presence of repeated eigenvalues, the non-uniqueness of eigenvectors causes the eigenvector derivatives undefined—in this case, an infinitesimal deviation from $p = 0$, can cause the eigenvectors to change from (17) to (16), discontinuously.

*Remark.* In this simple example, the discontinuity can be fixed by examining the limit of the $\mathbf{X}$ derivative as $p$ approaches 0, because in this case the derivative of $\mathbf{X}$ at $p \neq 0$ is well-defined. However, the situations encountered in many photonic design tasks can be much more challenging. For example, in Fig. 2 of the main text, not only are the eigenvalues (i.e., effective indices) repeated, their derivatives with respect to the parameter are also repeated (e.g, see mode 0 and mode 1 in Fig. 2-c). In those cases, the derivative of $\mathbf{X}$ is not well-defined any more. If up to the $n$-th order derivatives of the eigenvalues are repeated, one has to rely on the $n+1$-th order derivative of $\mathbf{X}$ to resolve the discontinuity. This is a rather expensive, if not impossible, computational process.

## 3 Derivation of Equations (5)

Here we present the derivation of Eqs. (5) in the main text. Starting from Eqs. (3), we first rewrite them as

$$\begin{aligned}
\mathbf{R}_\mathrm{L} = \mathbf{R}_\mathrm{R} &= \left[\mathbf{I} - \left(\mathbf{A}^{-1}\mathbf{X}\mathbf{B}\right)^2\right]^{-1} \mathbf{A}^{-1}\left(\mathbf{X}\mathbf{B}\mathbf{A}^{-1}\mathbf{X}\mathbf{A} - \mathbf{B}\right) \\
&= \left[\mathbf{I} - \left(\mathbf{A}^{-1}\mathbf{X}\mathbf{B}\right)^2\right]^{-1} \left(\mathbf{A}^{-1}\mathbf{X}\mathbf{B}\mathbf{A}^{-1}\mathbf{X}\mathbf{A} - \mathbf{A}^{-1}\mathbf{B}\right) \\
&= \left(\mathbf{I} - \mathbf{D}_1^2\right)^{-1} \left(\mathbf{D}_1\mathbf{D}_2 - \mathbf{D}_3\right),
\end{aligned} \quad (18)$$

and

$$\begin{aligned}
\mathbf{T}_\mathrm{LR} = \mathbf{T}_\mathrm{RL} &= \left(\mathbf{A} - \mathbf{X}\mathbf{B}\mathbf{A}^{-1}\mathbf{X}\mathbf{B}\right)^{-1} \mathbf{X}\left(\mathbf{A} - \mathbf{B}\mathbf{A}^{-1}\mathbf{B}\right) \\
&= \left[\mathbf{I} - \left(\mathbf{A}^{-1}\mathbf{X}\mathbf{B}\right)^2\right]^{-1} \mathbf{A}^{-1}\mathbf{X}\mathbf{A}\left[\mathbf{I} - \left(\mathbf{A}^{-1}\mathbf{B}\right)^2\right] \\
&= \left[\mathbf{I} - \left(\mathbf{A}^{-1}\mathbf{X}\mathbf{B}\right)^2\right]^{-1} \left(\mathbf{A}^{-1}\mathbf{X}\mathbf{A} - \mathbf{A}^{-1}\mathbf{X}\mathbf{B}\mathbf{A}^{-1}\mathbf{B}\right) \\
&= \left(\mathbf{I} - \mathbf{D}_1^2\right)^{-1} \left(\mathbf{D}_2 - \mathbf{D}_1\mathbf{D}_3\right).
\end{aligned} \quad (19)$$

In the above two expressions, the three matrix notations are defined as

$$\mathbf{D}_1 \coloneqq \mathbf{A}^{-1}\mathbf{X}\mathbf{B}, \ \mathbf{D}_2 \coloneqq \mathbf{A}^{-1}\mathbf{X}\mathbf{A}, \text{ and } \mathbf{D}_3 \coloneqq \mathbf{A}^{-1}\mathbf{B}. \quad (20)$$

To reach the expressions in Eqs. (6) of the main text, we first rewrite the expressions of $\mathbf{A}$ and $\mathbf{B}$ in Eqs. (4) as the following forms:

$$\begin{aligned}
\mathbf{A} &= \mathbf{W}^{-1}(\mathbf{W}_0 + \mathbf{W}\mathbf{V}^{-1}\mathbf{W}_0) \text{ and} \\
\mathbf{B} &= \mathbf{W}^{-1}(\mathbf{W}_0 - \mathbf{W}\mathbf{V}^{-1}\mathbf{W}_0).
\end{aligned} \quad (21)$$

Notice that $\mathbf{T}$, defined as $\mathbf{T} \coloneqq \mathbf{\Omega}\mathbf{Q}^{-1}$ in Sec. 3.B of the main text, has the equality relation $\mathbf{T} = \mathbf{W}\mathbf{V}^{-1}$. This is because, as introduced in Sec. 2, $\mathbf{V}$ is related to $\mathbf{W}$ through $\mathbf{V} = \mathbf{Q}\mathbf{W}\mathbf{\Lambda}^{-1}$, and thus we have

$$\mathbf{V} = \mathbf{Q}\mathbf{W}\mathbf{\Lambda}^{-1} = \mathbf{Q}\mathbf{\Omega}^{-1}\mathbf{\Omega}\mathbf{W}\mathbf{\Lambda}^{-1} = \mathbf{Q}\mathbf{\Omega}^{-1}\mathbf{W} = \mathbf{T}^{-1}\mathbf{W}, \quad (22)$$

where the third equality uses the fact that $\mathbf{W}$ is the eigenvector matrix of $\mathbf{\Omega}$. Substituting $\mathbf{T} = \mathbf{W}\mathbf{V}^{-1}$ into (21) and then in (20) yields the expressions in Eqs. (6) of the main text.

## 4 Derivative of Matrix Exponential

Provided the matrix $\mathbf{\Omega}$ and its derivative $\mathbf{\Omega}'$ with respect to a design parameter, we compute the derivative of the matrix exponential $e^{j\mathbf{\Omega}L/k_0}$ through another the matrix exponential $e^{j\mathbf{G}L/k_0}$, where $\mathbf{G}$ is defined in Eq. (10) in the main text, and we repeat it here for convenience:

$$\mathbf{G} = \begin{bmatrix} \mathbf{\Omega} & \mathbf{\Omega}' \\ \mathbf{0} & \mathbf{\Omega} \end{bmatrix}.$$

To compute this matrix exponential efficiently (and avoid eigenvalue decomposition), we tailor the scaling and squaring method [1] to leverage the particular matrix structure of $\mathbf{G}$. We refer to [1] for the details of scaling and squaring method. To present how we tailor it for efficient computation, we outline its major steps here:

1. Compute an integer $s$ to scale the matrix $j\mathbf{G}L/k_0$. Let $\tilde{\mathbf{G}}$ denote the scaled version: $\tilde{\mathbf{G}} = \frac{1}{2^s} j\mathbf{G}L/k_0$.

2. Compute two matrix polynomials, $D(\tilde{\mathbf{G}})$ and $N(\tilde{\mathbf{G}})$. The specific forms of these polynomials are determined by the scaling and squaring method.

3. Compute $\mathbf{A} = D^{-1}(\tilde{\mathbf{G}})N(\tilde{\mathbf{G}})$. The matrix inverse here is computed through matrix factorization.

4. Repeatedly square matrix $\mathbf{A}$ for $s$ times, resulting in $\mathbf{B} = \mathbf{A}^{2^s}$. $\mathbf{B}$ is a close approximation of $e^{j\mathbf{G}L/k_0}$.

We notice that the matrix $\mathbf{G}$ has a specific structure—its lower-left block matrix is always a zero matrix, and its two diagonal block matrices are the same. This structure is preserved in $\tilde{\mathbf{G}}$ as well as in its matrix polynomials. We leverage this structure for accelerating the computation in two aspects.

First, in the computation of a matrix polynomial (step 2), we need to compute some matrix powers $\tilde{\mathbf{G}}^n$ for some integers $n$. Typically, $\tilde{\mathbf{G}}^n$ is expressed recursively as $\tilde{\mathbf{G}}\tilde{\mathbf{G}}^{n-1}$, and each recursion requires a multiplication of two $N \times N$ matrices, where $N$ is the size of $\tilde{\mathbf{G}}$. In our computation, both $\tilde{\mathbf{G}}$ and $\tilde{\mathbf{G}}^{n-1}$ have the aforementioned block structure, and they can be represented as

$$\tilde{\mathbf{G}} = \begin{bmatrix} \mathbf{M}_1 & \mathbf{N}_1 \\ \mathbf{0} & \mathbf{M}_1 \end{bmatrix} \text{ and } \tilde{\mathbf{G}}^{n-1} = \begin{bmatrix} \mathbf{M}_2 & \mathbf{N}_2 \\ \mathbf{0} & \mathbf{M}_2 \end{bmatrix}. \tag{23}$$

Their product is

$$\tilde{\mathbf{G}}\tilde{\mathbf{G}}^{n-1} = \begin{bmatrix} \mathbf{M}_1\mathbf{M}_2 & \mathbf{M}_1\mathbf{N}_2 + \mathbf{N}_1\mathbf{M}_2 \\ \mathbf{0} & \mathbf{M}_1\mathbf{M}_2 \end{bmatrix}, \tag{24}$$

which involves three matrix multiplications of size $\frac{N}{2} \times \frac{N}{2}$. This is significantly faster than computing a multiplication of two $N \times N$ matrices.

Secondly, the two matrix polynomials $D(\tilde{\mathbf{G}})$ and $N(\tilde{\mathbf{G}})$ also have the same block structure, represented as

$$D(\tilde{\mathbf{G}}) = \begin{bmatrix} \mathbf{U}_1 & \mathbf{V}_1 \\ \mathbf{0} & \mathbf{U}_1 \end{bmatrix} \text{ and } N(\tilde{\mathbf{G}}) = \begin{bmatrix} \mathbf{U}_2 & \mathbf{V}_2 \\ \mathbf{0} & \mathbf{U}_2 \end{bmatrix}. \tag{25}$$

Then, then the matrix product $D^{-1}(\tilde{\mathbf{G}})N(\tilde{\mathbf{G}})$ (in step 3) can be expressed as

$$D^{-1}(\tilde{\mathbf{G}})N(\tilde{\mathbf{G}}) = \begin{bmatrix} \mathbf{U}_1^{-1}\mathbf{U}_2 & \mathbf{U}_1^{-1}\mathbf{V}_2 - \left(\mathbf{U}_1^{-1}\mathbf{V}_1\right)\left(\mathbf{U}_1^{-1}\mathbf{U}_2\right) \\ \mathbf{0} & \mathbf{U}_1^{-1}\mathbf{U}_2 \end{bmatrix},$$

which involves only the matrix inverse (factorization) of the $\frac{N}{2} \times \frac{N}{2}$ matrix $\mathbf{U}_1$ and a few matrix multiplications of size $\frac{N}{2} \times \frac{N}{2}$. This is also much faster than computing $D^{-1}(\tilde{\mathbf{G}})N(\tilde{\mathbf{G}})$ directly.


\end{document}